\documentclass[aps,nofootinbib,prd,preprint,preprintnumbers]{revtex4-1}
\usepackage{graphicx}
\usepackage{url}
\usepackage{xcolor}
\usepackage{amsmath}
\usepackage{amssymb}
\usepackage{hyperref}
\usepackage[normalem]{ulem}
\usepackage{soul}
\hypersetup{colorlinks=true,linkcolor=redLinks,citecolor=greenLinks,
urlcolor=redLinks,
pdfborder={0 0 1}}
\hypersetup{
    bookmarks=true,         
    unicode=false,          
    pdftoolbar=true,        
    pdfmenubar=true,        
    pdffitwindow=false,     
    pdfstartview={FitH},    
    pdftitle={My title},    
    pdfauthor={Author},     
    pdfsubject={Subject},   
    pdfcreator={Creator},   
    pdfproducer={Producer}, 
    pdfkeywords={keyword1, key2, key3}, 
    pdfnewwindow=true,      
    colorlinks=false,       
    linkcolor=red,          
    citecolor=green,        
    filecolor=magenta,      
    urlcolor=cyan           
}
\colorlet{shadecolor}{gray!15}
\definecolor{greenLinks}{rgb}{0, 0.6, 0} 
\definecolor{YELLOW}{rgb}{0.5, 0, 0.5} 
\definecolor{blueLinks}{rgb}{0, 0, 0.6}
\definecolor{redLinks}{rgb}{0.6, 0, 0}
\definecolor{tempText}{rgb}{0.55, 0.10,0.67}
\definecolor{eprintLinks}{rgb}{0.4, 0.4, 0.4}
\definecolor{journalLinks}{rgb}{0.6, 0, 0}

\usepackage[T1]{fontenc} 
\usepackage{hyperref}
\usepackage{float}
\usepackage{mathtools}

\def\21{$\mathrm{SU(2)_L \otimes U(1)_Y}$ }

\newcommand{\AddrAHEP}{AHEP Group, Institut de F\'{i}sica Corpuscular --
  C.S.I.C./Universitat de Val\`{e}ncia, Parc Cientific de Paterna.\\
  C/Catedratico Jos\'e Beltr\'an, 2 E-46980 Paterna (Val\`{e}ncia) - SPAIN}

\definecolor{darkgreen}{RGB}{60,179,113}
\definecolor{darkyellow}{RGB}{153,153, 0}

\definecolor{gesfpurple}{rgb}{0.47,0.19,0.42}

\definecolor{gesflanse}{rgb}{0.00,0.50,0.50}

\definecolor{gesfblue}{rgb}{0.08,0.42,0.76}

\definecolor{gesfred}{rgb}{1,0,0}

\definecolor{gesfwhite}{rgb}{1,1,1}

\definecolor{gesfblack}{rgb}{0,0,0}

\def\31{$\mathrm{SU(3)_c \otimes U(1)_Q}$ }

\newcommand{\beq}{\begin{equation}}
\newcommand{\eeq}{\end{equation}}                     
\newcommand{\beqa}{\begin{eqnarray}}
\newcommand{\eeqa}{\end{eqnarray}}          

\newcommand {\ignore}[1]{}

\begin{document}

  \title{Neutrino oscillations from warped flavor symmetry:\\
   predictions for long baseline experiments  T2K, NOvA and DUNE}

\author{Pedro Pasquini~$^{1,2}$}\email{pasquini@ifi.unicamp.br}
\author{S. C. Chuli\'a}\email{salcen@alumni.uv.es}
\author{J. W. F. Valle~$^2$} \email{valle@ific.uv.es, URL:
  http://astroparticles.es/} 
\affiliation{$^1$~Instituto de F\'isica Gleb Wataghin - UNICAMP, {13083-859}, Campinas SP, Brazil}
\affiliation{$^2$~\AddrAHEP}

\begin{abstract}
  
  Here we study the pattern of neutrino oscillations emerging from a
  previously proposed warped model construction incorporating
  $\Delta(27)$ flavor symmetry~\cite{Chen:2015jta}.
  In addition to a complete description of fermion masses, the model
  predicts the lepton mixing matrix in terms of two parameters.
  The good measurement of $\theta_{13}$ makes these two parameters
  nearly proportional, leading to an approximate one-parameter
  description of neutrino oscillations.
  There is a sharp fourfold degenerate correlation between
  $\delta_{CP}$ and the atmospheric mixing angle $\theta_{23}$, so
  that maximal $\theta_{23}$ also implies maximal leptonic CP
  violation.
  The predicted electron neutrino and anti-neutrino appearance
  probabilities indicate that the model should be tested at the T2K,
  NO$\nu$A and DUNE long baseline oscillation experiments.
  
 \end{abstract}


 \maketitle

 \section{Introduction}

 The    striking    pattern    of    neutrino    mass    and    mixing
 parameters~\cite{Agashe:2014kda},  remarkably   at  odds   with  that
 characterizing  the quark  sector,  suggests that  it  can hardly  be
 expected to happen just by chance.
 By and large, in the attempt to bring a rationale to the pattern of
 fermion mixing, theorists have focused on the idea that there is
 some yet-to-be-determined non-Abelian flavor symmetry of
 nature.
 Theoretical work towards predicting flavor parameters has followed
 two complementary paths, namely:
 \begin{itemize}
 \item Building explicit flavor models on a case-by-case
   basis~\cite{babu:2002dz,Altarelli:2010gt,Morisi:2012fg,Morisi:2013qna,King:2014nza}
 \item Focussing upon the residual CP symmetries characterizing the
   final mass matrices, irrespective of the details of the underlying
   theory~\cite{Chen:2015siy,Chen:2016ica}
 \end{itemize}
 In both cases a number of predictions for the leptonic mixing matrix
 can be made. Specially interesting for phenomenology are those which
 imply a correlation between the leptonic Jarlskog invariant
 $$J_{\rm CP}={\rm Im}[U_{e1}^*U_{\mu 3}^*U_{\mu 1}U_{e3}]$$
 and the atmospheric angle, since $\delta_{\rm CP}$ and $\theta_{23}$
 are the two less precisely determined of all the neutrino oscillation
 parameters.

 Despite enormous experimental effort since the discovery of neutrino
 oscillations~\cite{Kajita:2016cak,McDonald:2016ixn}, we are still far
 from a high precision measurement of the leptonic CP phase
 characterizing neutrino oscillations within the three flavor
 paradigm~\cite{Forero:2014bxa}.
 In this paper we focus on a model where neutrino oscillations are, to
 a good approximation, described in terms of a unique parameter that
 may be taken as $\delta_{\rm CP}$.

 For definiteness, we focus upon symmetry based flavor models.  We
 consider the case of a full-fledged warped model construction
 incorporating flavor symmetry~\cite{Chen:2015jta}. It provides an
 explicit proof-of-concept, realizing a realistic and ``natural''
 scheme providing a complete description of all fermion masses as well
 as their mixing angles and phases.
 All mass hierarchies are naturally accounted for by warping. Thanks
 to the underlying $\Delta(27)$ flavor symmetry, the model implies a
 predictive pattern of lepton mixing parameters, while adequately
 fitting the quark mixing matrix.
 Here we determine the implications of the model for neutrino
 oscillations, deriving the manifest correlation between $\delta_{CP}$
 and the atmospheric angle $\theta_{23}$.
 We find that the determination of leptonic CP violation is no longer
 unique but exhibits a fourfold degeneracy.
 This is used to work out the predicted electron neutrino and
 anti-neutrino appearance probabilities for the long baseline neutrino
 oscillation experiments T2K, NOvA and the upcoming
 DUNE~\cite{Acciarri:2015uup} experiment.

 \section{Theory preliminaries: the warped flavor model}
 
  Here the main idea consists in combining the advantages of warping in
 order to account for mass hierarchies without fine tuning, with those
 of implementing flavor symmetries, in order to potentially predict the
 fermion mixing pattern.
 For definiteness we focus on the Warped Flavor Model proposed
 in~\cite{Chen:2015jta}. The model is a minimal version of the
 Randal-Sundrun (RS) model involving compactification of the fifth
 dimension on $S_1/{\cal Z}_2$ and attaching the orbifold to $y=0$ (UV
 brane) and $y=L$ (IR brane).
 The warped five-dimensional $AdS_5$ metric is given as
\begin{equation}
ds^2=e^{-2ky}\eta_{\mu\nu}dx^\mu dx^\nu-dy^2,
\end{equation}
where $\eta_{\mu\nu}={\rm Diag}[1,-1-1-1]$ and $k$ is the curvature
scale parameter.

  We employ the most minimal version of the RS model with
  non-custodial $G_{\rm SM}=SU(2)_{\rm L}\otimes U(1)_{\rm Y}$ bulk
  eletroweak symmetry where the 5D fermions and the Higgs field, $H$,
  are allowed to propagate into the bulk.
  Although models with a brane-localized Higgs and no custodial
  symmetry are severely constrained by electroweak precision
  tests~\cite{Csaki:2002gy} the conflict with electroweak and flavour
  physics constraints can be significantly reduced when the 5D Higgs
  field lives in the
  bulk~\cite{Cabrer:2011fb,Carmona:2011ib,Archer:2014jca,Agashe:2008uz,Archer:2011bk,Cabrer:2011qb}.
  This offers a way to account for fermion mass hierarchies while
  evading eletroweak and flavour physics restrictions.
  Besides the usual SM states, 4 extra scalars are added to the model,
  $\phi,\sigma_1,\sigma_2$ and $\xi$, all of which acquire vacuum
  expectation values.
  It has been shown that charged lepton as well as Dirac neutrino
  masses are generated at leading order~\cite{Chen:2015jta}.
  Moreover, all fermion mass hierarchies can be adequately described
  by appropriate choices of bulk mass parameters~\cite{Chen:2015jta}.
  On the other hand {\it Dirac} neutrino masses are generated by
  interactions in the IR brane of the form
  $\sim (\xi\sigma_a\overline{\Psi}_l)\tilde{H}\Psi_{\nu_i}$. As a
  result, the double vev of the scalars and the localization of the
  interaction can naturally account for their smallness.
   
  Concerning the mixing angles, a beautiful feature of the model
  consists in the integration of its extra-dimensional nature with
  the implementation of a non-Abelian flavor symmetry, in our case
  $\Delta(27)\otimes {\cal Z}_4\otimes {\cal Z}'_4$.
  The latter leads to the prediction of all the 4 neutrino oscillation
  parameters in terms of just two angles: $\theta_\nu$ and $\phi_\nu$
 according to the following equations,
\begin{flalign}\label{eq:predict1}
\sin^2\theta_{12}=&\frac{1}{2-\sin2\theta_\nu\cos\phi_\nu} \\\label{eq:predict2}
\sin^2\theta_{13}=&\frac{1}{3}(1+\sin2\theta_\nu\cos\phi_\nu) \\\label{eq:predict3}
\sin^2\theta_{23}=&\frac{1-\sin2\theta_\nu\sin(\pi/6-\phi_\nu)}{2-\sin2\theta_\nu\cos\phi_\nu}\\ \label{eq:predict4}
J_{\rm CP}=&-\frac{1}{6\sqrt{3}}\cos2\theta_\nu,
\end{flalign}
where $J_{\rm CP}$ is the Jarlskog invariant.

\begin{figure}[!h]
 \centering
  \includegraphics[scale=0.46,height=5.5cm]{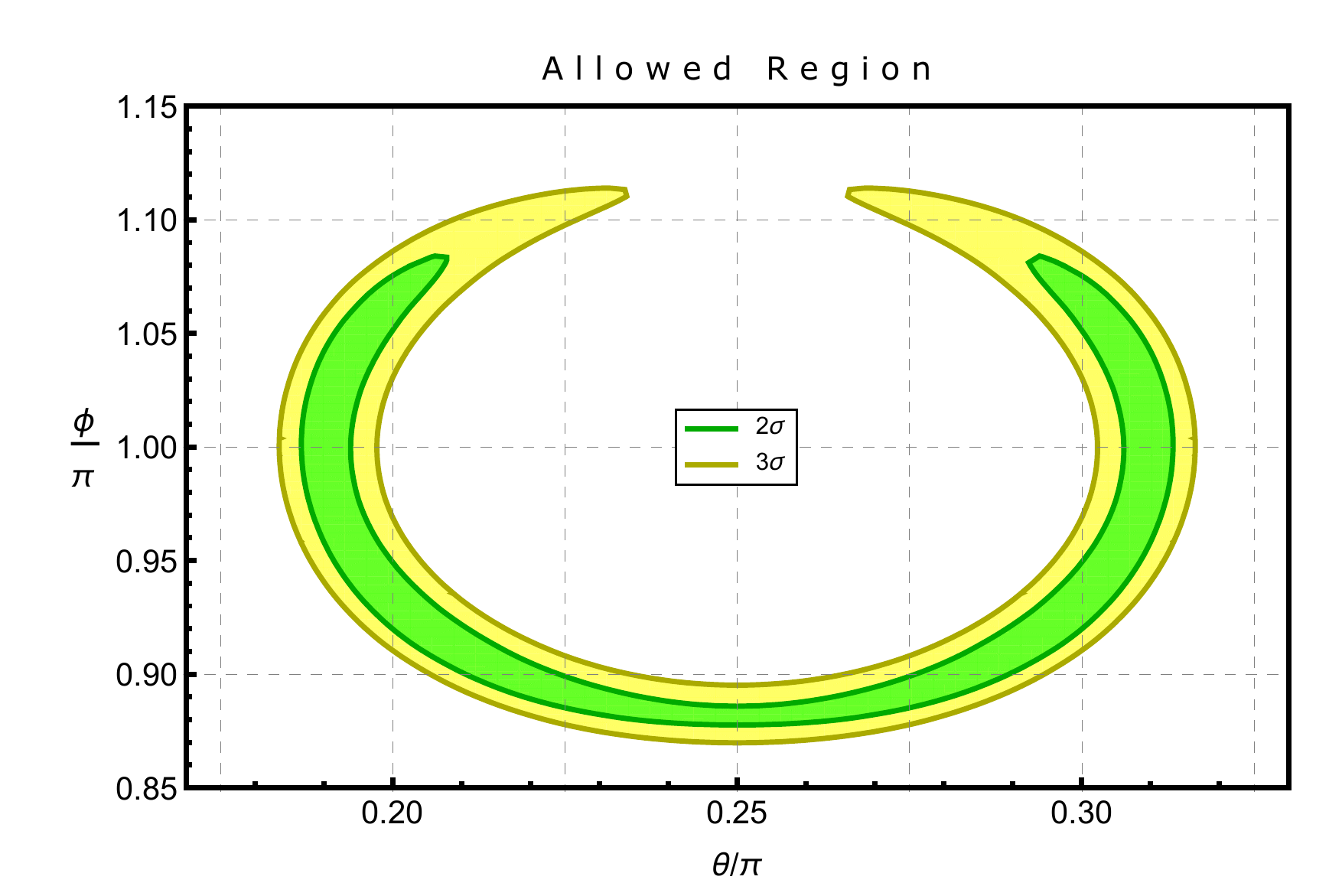}
  \includegraphics[scale=0.46,height=5.5cm]{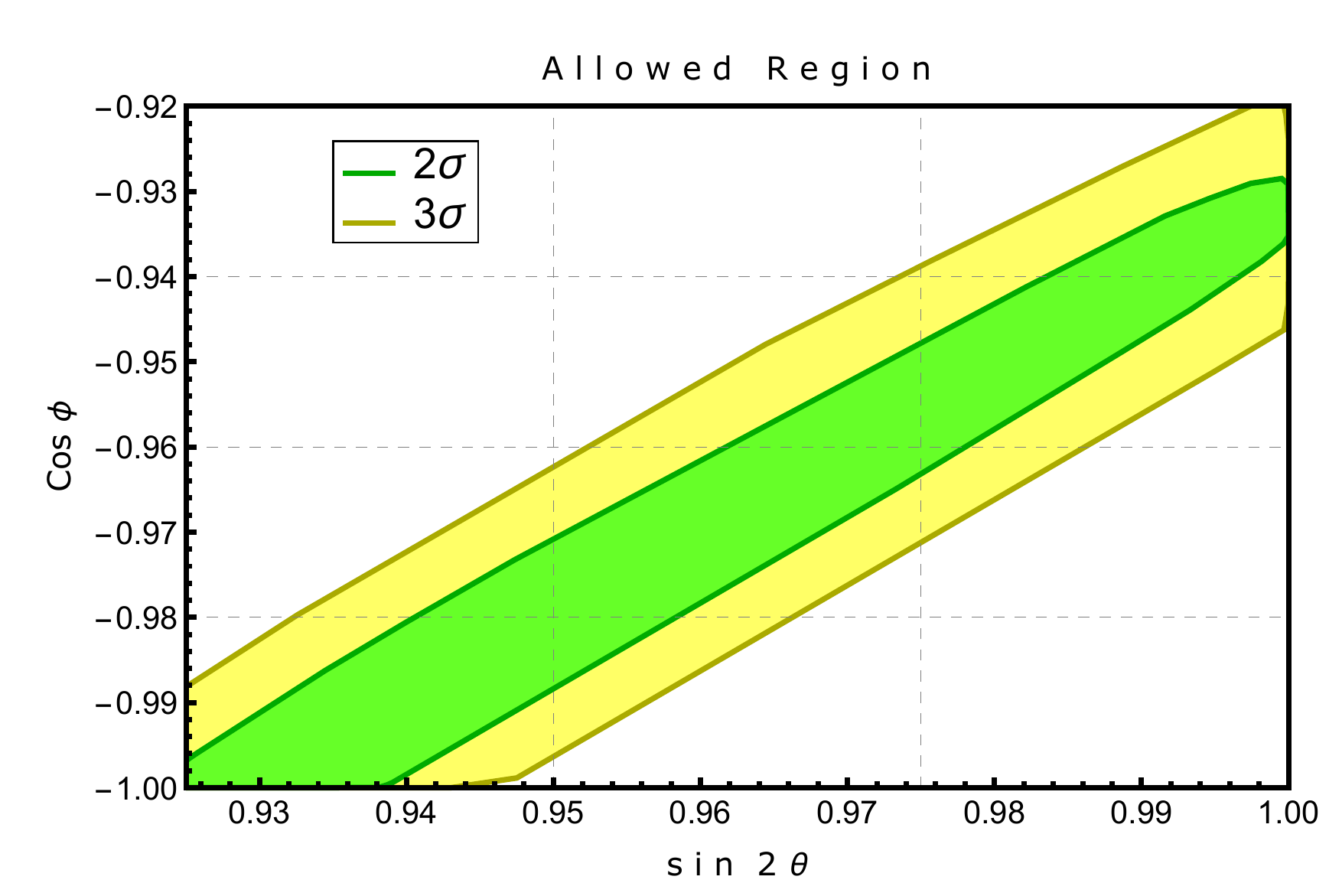}
  \caption{Two ways of displaying the $3\sigma$ (yellow) and $2\sigma$
    (green) regions of model parameters $\theta_\nu$ and $\phi_\nu$
    allowed by global fit of current neutrino oscillation data.}
    \label{fig:range}
  \end{figure}

These relations were derived in~\cite{Chen:2015jta} and they imply
that the theory can be directly probed using low energy neutrino
oscillation experiments by comparing the above predicted relations,
given in terms of the two free parameters $\theta_\nu$ and $\phi_\nu$,
with the measured oscillation parameters.
Indeed, the allowed parameter region consistent with current
oscillation data~\cite{Forero:2014bxa} can be determined and was
given in \cite{Chen:2015jta}. Here we refine and extend the analysis
so as to isolate the implications of the model. In
Fig.~\ref{fig:range} we present the $3\sigma$ and $2\sigma$ regions of
model parameters allowed by the current neutrino oscillation global
fit in \cite{Forero:2014bxa}.

   In the left panel of this figure we show the $3 \sigma$ and
    $2 \sigma$ allowed contours obtained using the global $\chi^2$
    function,
\begin{equation}
\chi^2_{\rm Global}=\chi_{12}^2+\chi_{13}^2+\chi_{23}^2,
\end{equation}
where in $\chi^2_{1i}$ we assumed Gaussian prior corresponding to the
global fit determinations of $\sin^2\theta_{12}$ and
$\sin^2\theta_{13}$,
\begin{equation}
\chi^2_{1i}=\left(\frac{\sin^2\theta_{1i}^{\rm exp}-\sin^2\theta_{1i}^{\rm pred}}{\sigma_{1i}}\right)^2
\end{equation}
for $i=2,3$, the $\sigma_{1i}$ the Gaussian error at $1\sigma$, while
the $\chi^2_{23}$ is the function for $\sin^2\theta_{23}$ used
in~\cite{Forero:2014bxa}.  Following~\cite{Chen:2015jta}, the
predicted values for the mixing angles $\theta_{ij}$ are calculated by
varying $\theta_\nu$ and $\phi_\nu$ given in
Eqs.~\ref{eq:predict1}-\ref{eq:predict3}~\footnote{Note however that,
  instead of taking the individual $\chi^2$ as in~\cite{Chen:2015jta},
  here we add them up, resulting in a slightly more restrictive
  constraint, as expected.}, leading to the regions depicted in the
left panel in Fig.~\ref{fig:range}. In the right panel we highlight
the effect of the precise determination of the reactor mixing angle
$\theta_{13}$ at Daya Bay.  Indeed, from Eq.~\ref{eq:predict2} it
follows that the two model parameters $\theta_\nu$ and $\phi_\nu$ are
sharply correlated, as displayed in the figure. Hence, in an
approximate sense our model is an effectively one parameter model for
neutrino oscillations.

\begin{figure}[!h]
 \centering
  \includegraphics[scale=0.8]{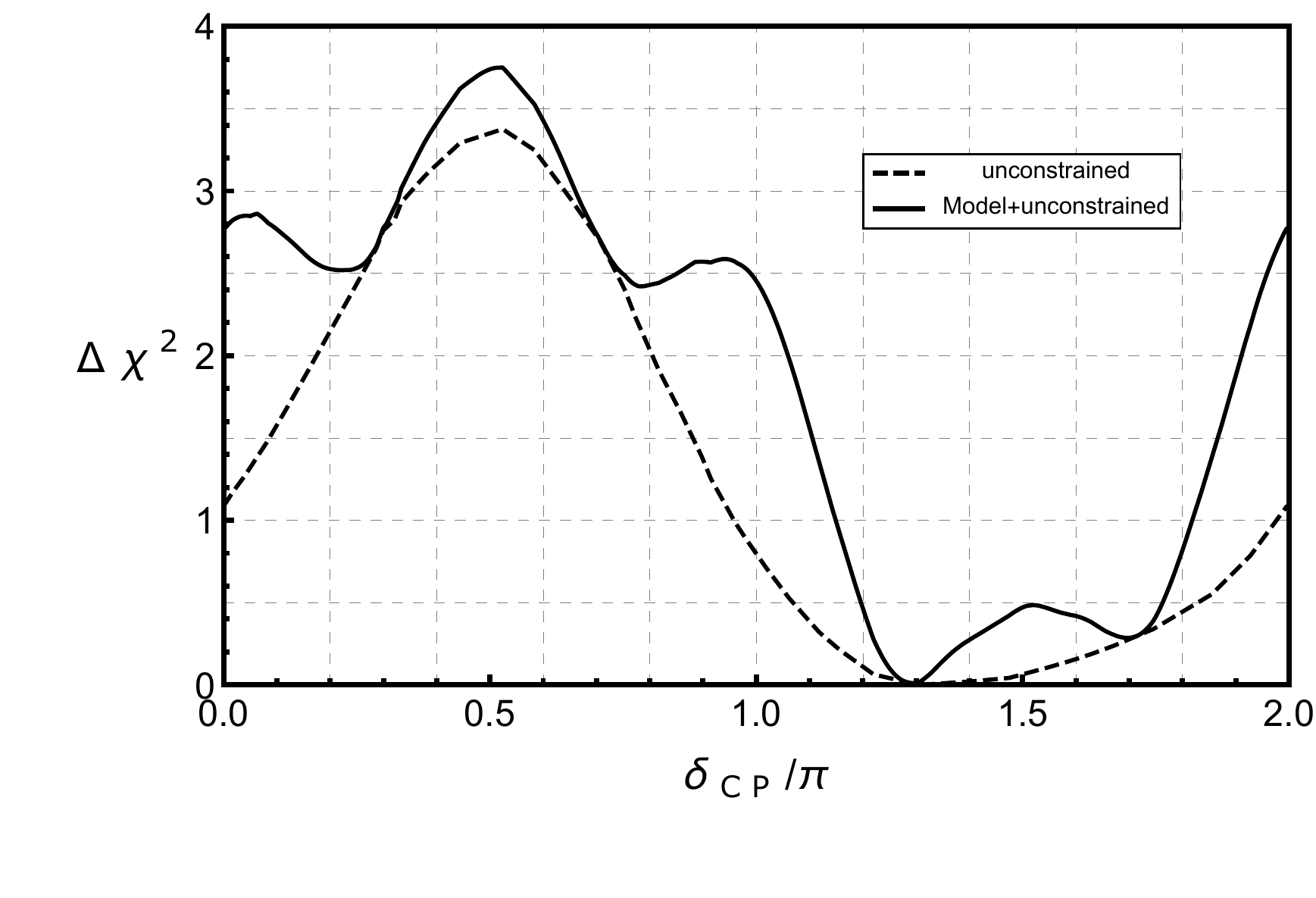}
  \caption{$\Delta \chi^2$ as a function of $\delta_{\rm CP}$ for our
    model (solid line) and for the general case (dashed line).}
    \label{fig:chi_delta}
  \end{figure}

\begin{figure}[!h]
 \centering
    \includegraphics[scale=0.46]{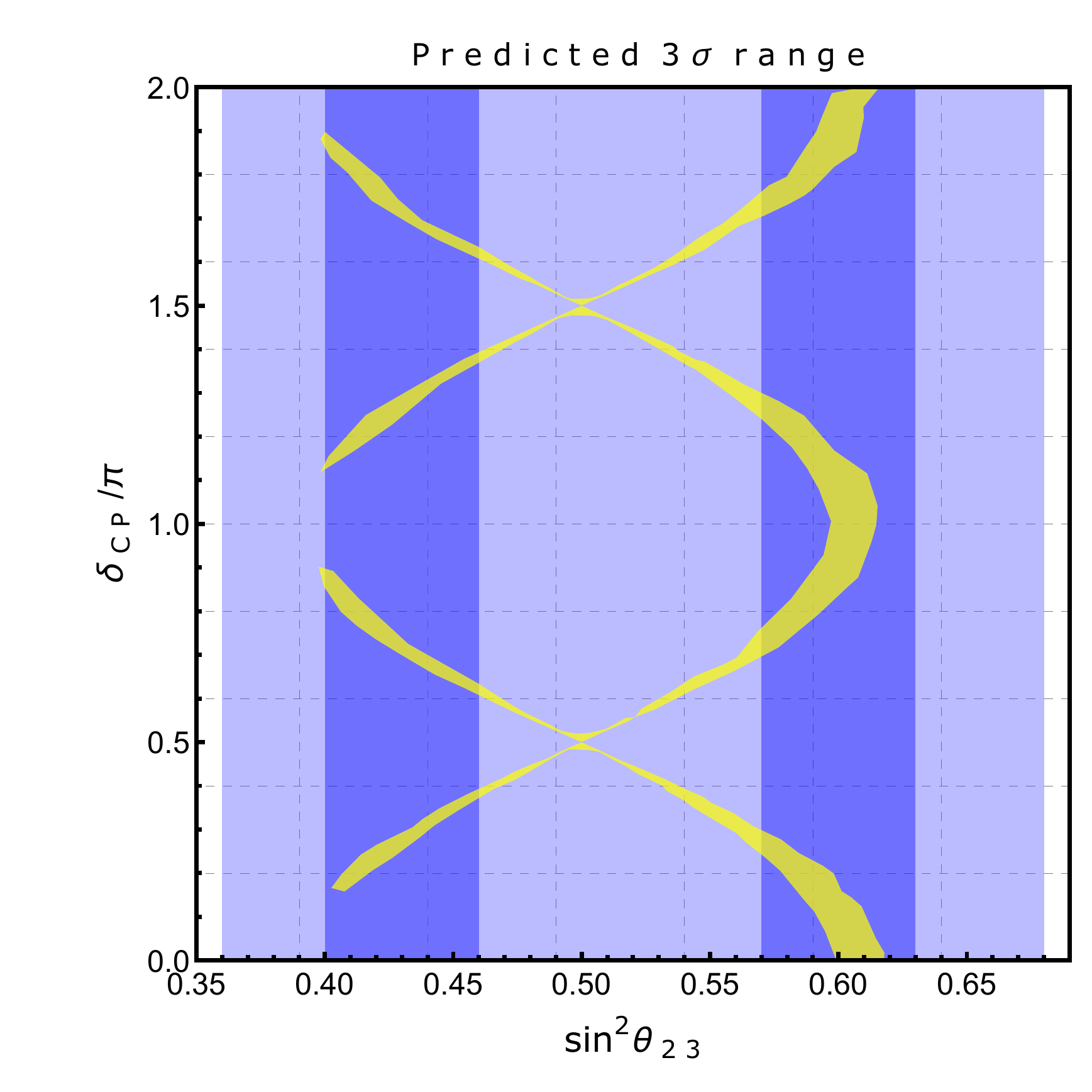}
    \includegraphics[scale=0.46]{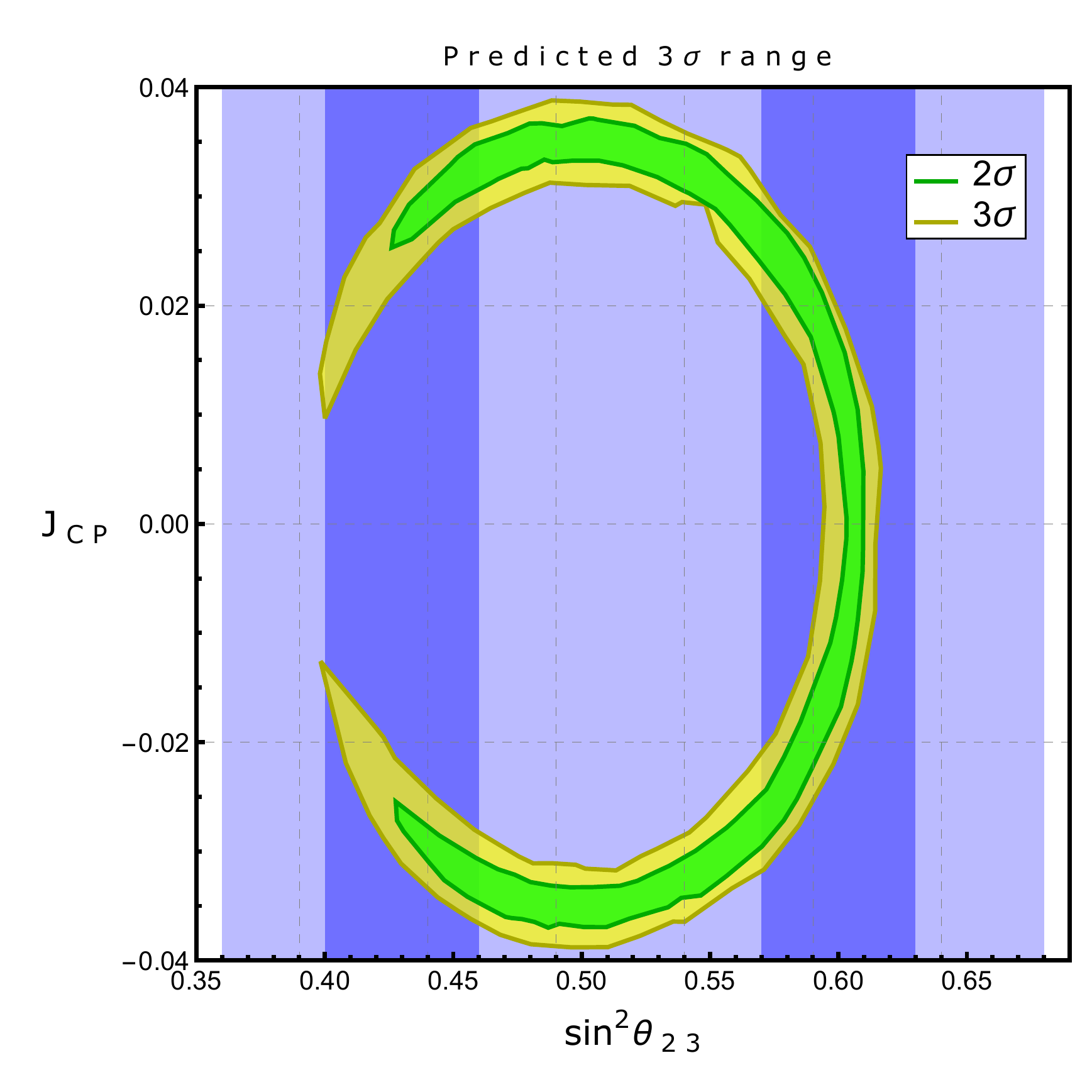}
    
    \caption{ Values of leptonic CP violation parameter allowed by
      current oscillation data versus the atmospheric angle
      $\theta_{23}$. Vertical bands are allowed at $3\sigma$ (light
      blue) and $1\sigma$ (darker blue) by the global fit
      in~\cite{Forero:2014bxa}. {\bf Left:} The yellow region denotes
      the $3\sigma$ region of $\delta_{\rm CP}$.  {\bf Right:} The
      yellow and green regions denote the $3\sigma$ and $2\sigma$
      regions of $J_{\rm CP}$, the Jarlskog invariant.  }
  \label{fig:correlation}
  \end{figure}

The Global $\chi^2$-analysis has two minima $\chi^2_{\rm min}=3.81$
found for: (1) $\theta_\nu^1=0.295\pi$ and $\phi_\nu=0.92\pi$ and (2)
$\theta_\nu^2=0.205\pi$ and $\phi_\nu=0.92\pi$ corresponding to the
mixing parameters given in Table~\ref{tab:predic}. This degeneracy is
trigonometric, as $\theta_{\nu}^i$ are complementary, that is,
$\theta_{\nu}^{1}+\theta_{\nu}^{2}=90^\circ$ and
Eqs.~\ref{eq:predict1}-\ref{eq:predict3} contain only
$\sin2\theta_\nu$. However this degeneracy is lifted by a measurement
of the Jarlskog invariant.  The latter is proportional to
$\cos2\theta_\nu$ and, by itself, gives rise to a fourfold degeneracy
in $\delta_{\rm CP}$ as can be seen in
Fig.\ref{fig:correlation}. Together with the neutrino oscillation data
determining the mixing angles one finds a solution
$\delta_{\rm CP}=1.27\pi$ deeper than the other minima, as presented
in Fig.\ref{fig:chi_delta}.
  \begin{table}[H]
  \centering
  \begin{tabular}{cccc}
  \hline \hline 
  Parameter & Minimum 1 & Minimum 2 & Unconstrained case\\\hline \hline
  $s_{12}^2$& $0.341$ &$0.341$ & 0.323($\pm$0.016)\\
  $s_{13}^2$ &$0.0232$ &$0.0232$ & 0.0234($\pm$0.0020) \\
  $s_{23}^2$ & $0.570$ &$0.570$ & 0.573($\substack{+0.025 \\ -0.043}$)\\
  $\delta_{\rm CP}/\pi$ & $1.27 (1.72)$ & $0.72 (0.27)$ & 1.34($\substack{+0.64 \\ -0.38}$)\\ \hline
  \end{tabular}
  \caption{\label{tab:predic} Predicted values of the neutrino
    oscillation parameters $s_{ij}^2=\sin^2\theta_{ij}$ corresponding
    to the $\chi^2$ minima obtained in our model for different
    ($\theta_\nu, \phi_\nu$) values.
    The fourth column denotes the standard ``unconstrained''
    three-neutrino best fit values from~\cite{Forero:2014bxa}.  }
\end{table}  
One notices that the preferred value of $\delta_{\rm CP}$ in our model
can be quite different from that found in the general
``unconstrained'' three-neutrino oscillation scenario. To see this
closer we plot in Fig.~\ref{fig:chi_delta} the $\Delta \chi^2$
function assuming our model to be true and compare with what is found
in the general ``unconstrained'' three-neutrino oscillation global
fit.
  The solid black curve gives the total $\Delta \chi^2$ in our model
  as $\chi^2_{\rm Total}=\chi^2_{\rm Global}+\chi^2_{\delta}$,
  while the black-dashed one corresponds to the ``unconstrained''
  $\chi^2_{\delta}$ function in~\cite{Forero:2014bxa}.
One notices that there are four nearly degenerate preferred
minima of $\delta_{\rm CP}$ in our model, all allowed at
$2\sigma$. These minima are narrower than the unique determination of
$\delta_{\rm CP}$ found in the general ``unconstrained'' case.
By providing an improved determination of $\delta_{\rm CP}$, future
data could make the difference between our model and the general case
potentially significant. Namely, within our model $\delta_{\rm CP}$
has a more precisely determined value than in the general case.

In addition, an improved $\delta_{\rm CP}$ determination would also
imply a determination of the atmospheric mixing angle $\theta_{23}$ in
our model, as seen in Fig.~\ref{fig:correlation} below.
Indeed our model predicts a sharp correlation between
$\delta_{\rm CP}$ and $\theta_{23}$ as illustrated in
Fig.\ref{fig:correlation}.
  As one can see, a feature of this predicted correlation is that 
  maximal mixing $\theta_{23}=\pi/4$ also implies maximal CP
  violation (up to sign), a remarkable prediction indeed.

  \section{Predictions for the T2K, NOvA and DUNE experiments}\label{sec:prob}

  As we saw above, the fact that the model contains only two tightly
  correlated free parameters suggests that it will be tested in future
  long baseline neutrino oscillation experiments.
  In order to see the effect of the restricted neutrino oscillation
  parameter space we have mapped out the resulting allowed values for
  the oscillation probability and compared with those expected in a
  generic model.
  To do this we varied the parameters $\theta_\nu$ and $\phi_\nu$
  inside the $3\sigma$ range of Fig.~\ref{fig:range} (Left). Of
  special interest here are those experiments that can probe the
  parameter $\delta_{\rm CP}$ (or $\theta_{13}$), as these play a key
  role to constrain our parameter space. Indeed, any of these can be
  taken as ``the'' key parameter of our model.

  The presence of CP violation in long baseline accelerator neutrino
  oscillation experiments would be manifest in the most direct way
  through the non-vanishing of the neutrino oscillation CP asymmetry
  $A_{\mu e}=(P_{\mu e}-P_{\overline{\mu}~ \overline{e}})/(P_{\mu
    e}+P_{\overline{\mu}~\overline{e}}).$ However here we prefer to
  display the individual neutrino and antineutrino oscillation
  probabilities, as these contain all the information. Matter effects
  are included in our calculations according to
  Ref.~\cite{Nunokawa2008338}.
  
\begin{figure}[!h]  
 \centering
  \includegraphics[scale=0.4]{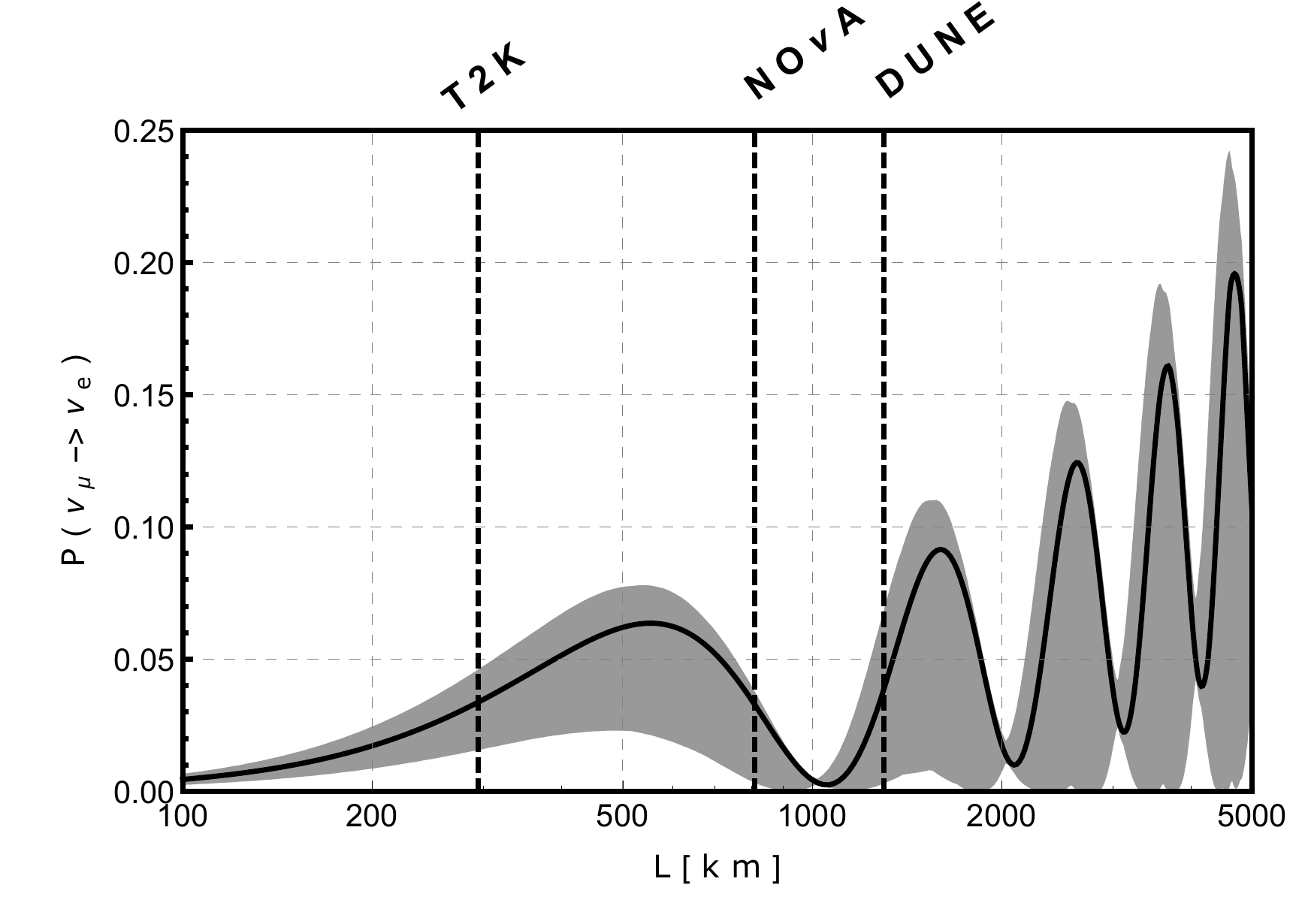}
    \includegraphics[scale=0.46]{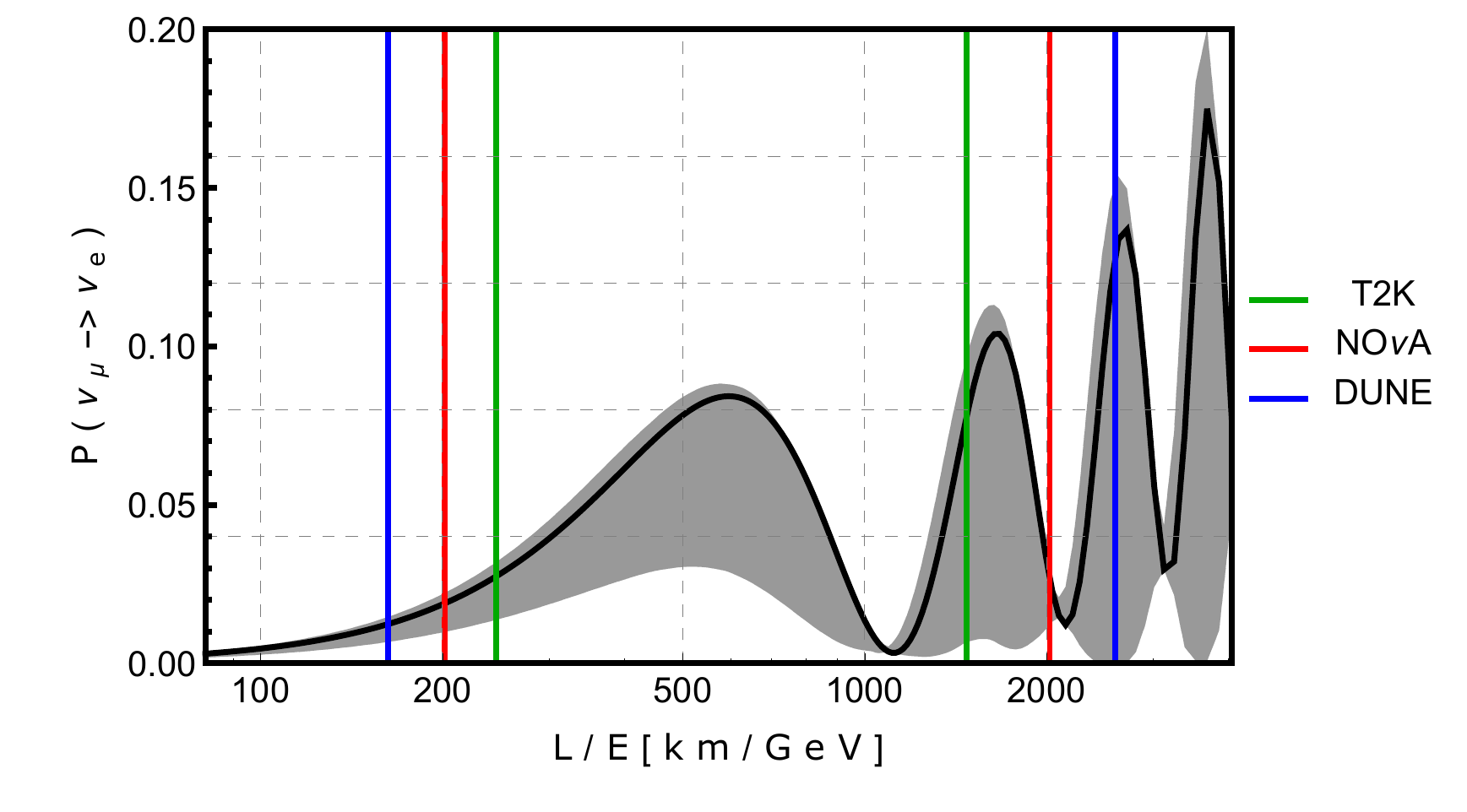}
    \caption{\label{fig:Loscilation} {\bf Left:} Transition
      probability $P(\mu\to e)$ as a function of the distance, $L$, of
      the detector for a $1$ GeV neutrino expected in the generic
      three-neutrino oscillation scheme.  The shaded region represents
      the $3\sigma$ range while the solid line corresponds to the
      current best fit. Vertical lines indicate the T2K, NOvA and
      DUNE baselines. {\bf Right:} The probability as a function of
      $L/E$, the color vertical lines indicating the $L/E$ ranges
      covered in each experiment.  } 
  \end{figure}

  In the left panel in Fig.~\ref{fig:Loscilation} we show the allowed
  regions of the oscillation probability expected within the generic
  three-neutrino oscillation scheme, as a function of the detector
  distance for a 1 GeV neutrino.
  The large spread in the allowed region for the appearance
  probabilities follows mainly from our poor knowledge of the
  $\delta_{\rm CP}$ phase.

\begin{figure}[H]
 \centering
  \includegraphics[scale=0.46]{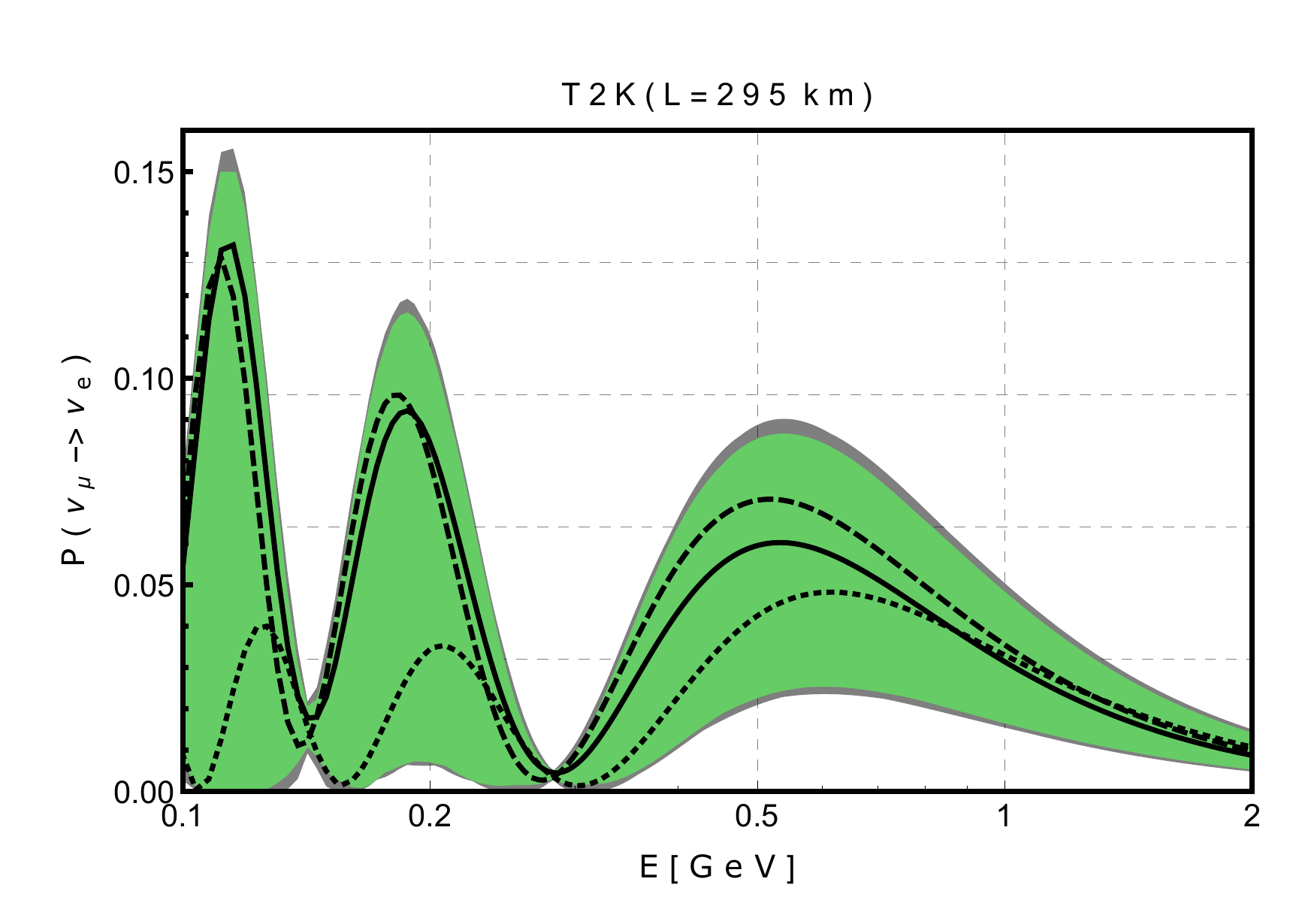}
    \includegraphics[scale=0.46]{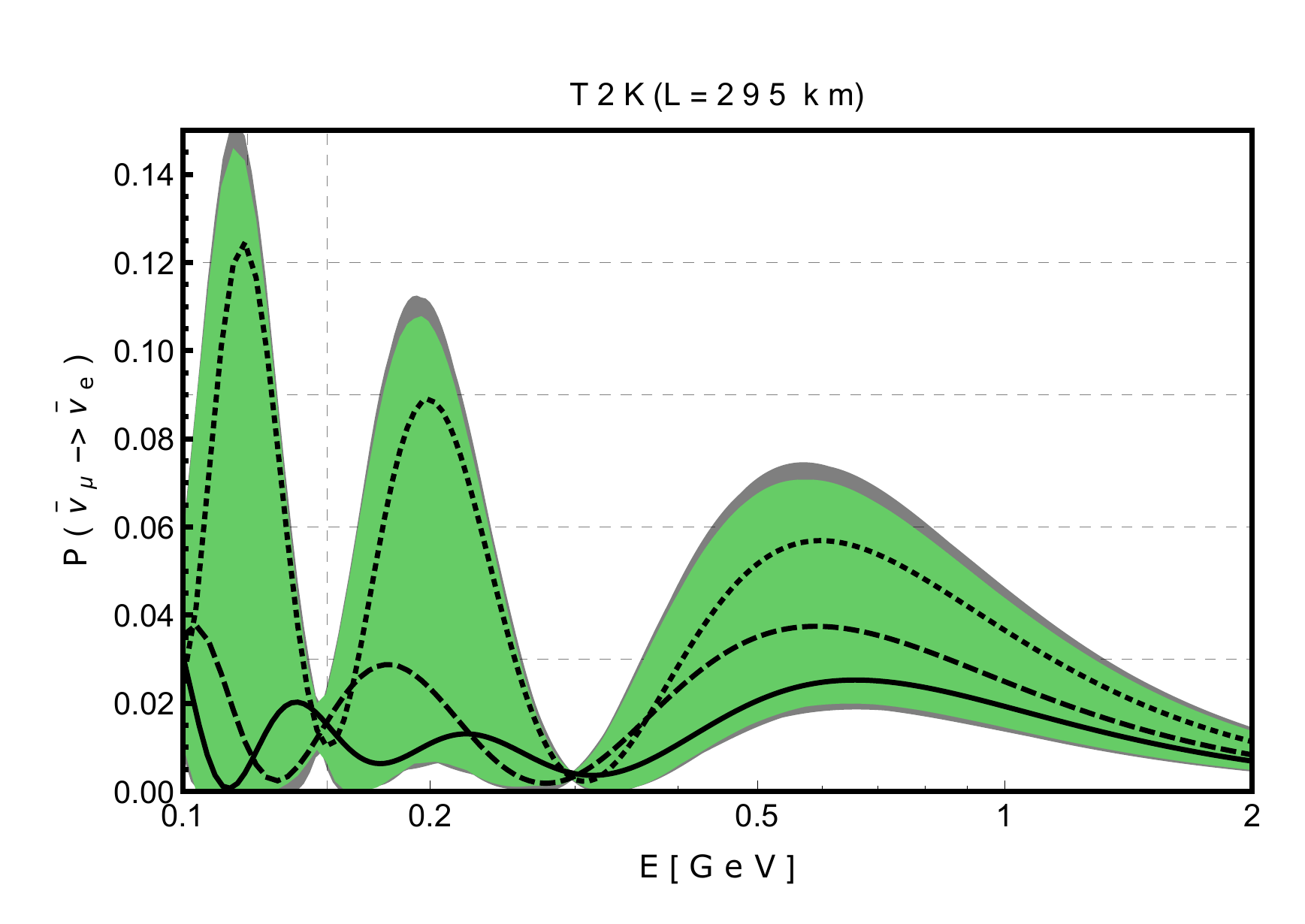}
    \caption{\label{fig:T2K} Transition probability as a function of
      the neutrino energy, $E$, for the T2K experiment. The green area
      represents the 3$\sigma$ allowed region in the model while the
      gray one represents the 3$\sigma$ allowed region in the
      unconstrained case. The full line represents the current
      neutrino best fit with $\delta_{CP}=3\pi/2$, the dashed line is
      the minimum 1 for $\delta_{CP}=1.27\pi$ and the dotted line the
      minimum 2 for $\delta_{CP}=0.27\pi$. {\bf Left:} $P(\mu\to e)$
      transition probability. {\bf Right:} $ P(\bar{\mu}\to \bar{e})$
      transition probability.}
  \end{figure}

\begin{figure}[!h]
 \centering
  \includegraphics[scale=0.46]{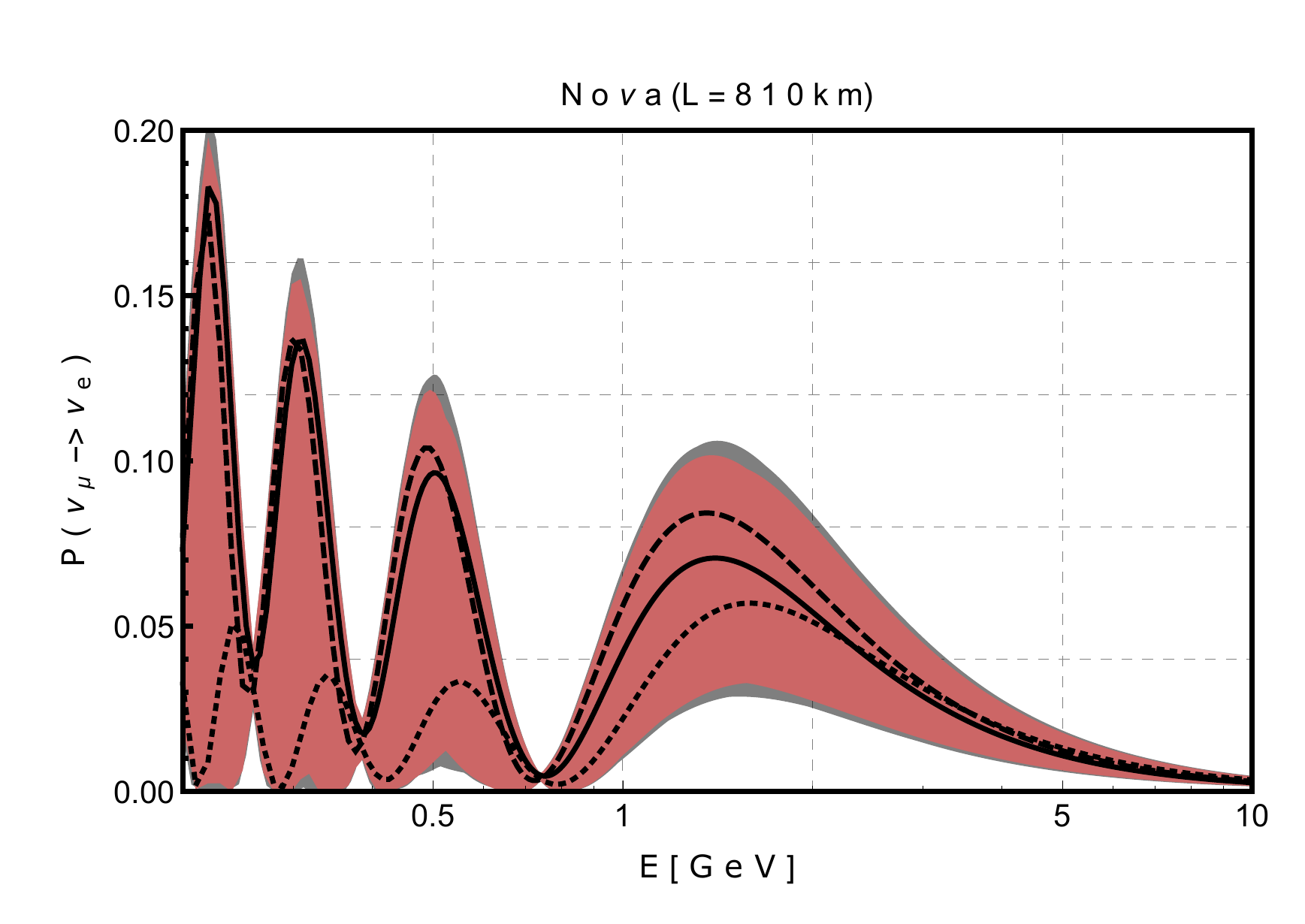}
    \includegraphics[scale=0.46]{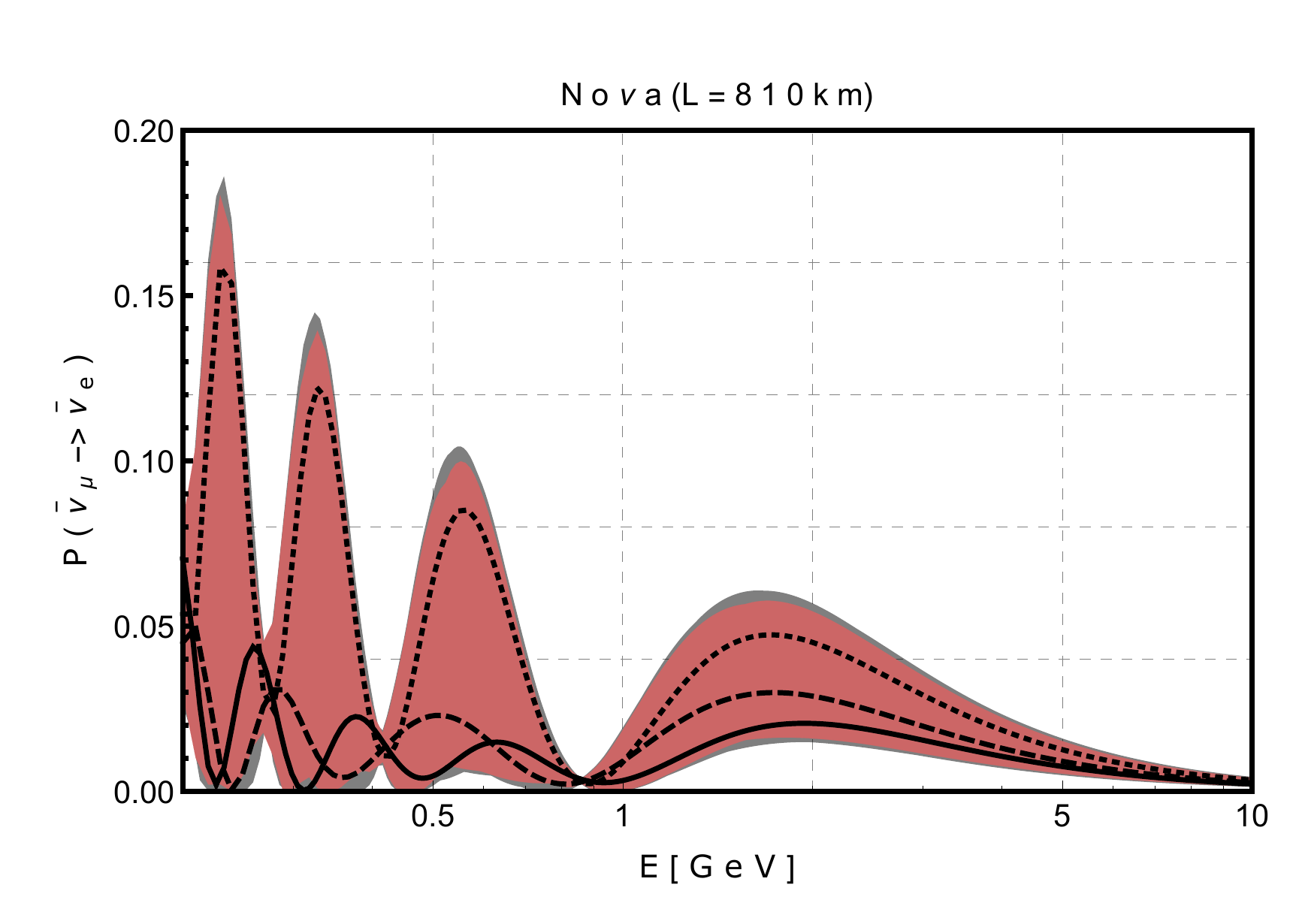}
    \caption{\label{fig:Nova} Transition probability as a function of
      the neutrino energy, $E$, for the No$\nu$a experiment.  The red
      area represents the 3$\sigma$ allowed region in the model while
      the gray one represents the 3$\sigma$ allowed region in the
      unconstrained case. The full line represents the current
      neutrino best fit with $\delta_{CP}=3\pi/2$, the dashed line is
      the minimum 1 for $\delta_{CP}=1.27\pi$ and the dotted line the
      minimum 2 for $\delta_{CP}=0.27\pi$. {\bf Left:} $P(\mu\to e)$
      transition probability. {\bf Right:} $ P(\bar{\mu}\to \bar{e})$
      transition probability.}
 \end{figure}

\begin{figure}[!h]
 \centering
  \includegraphics[scale=0.46]{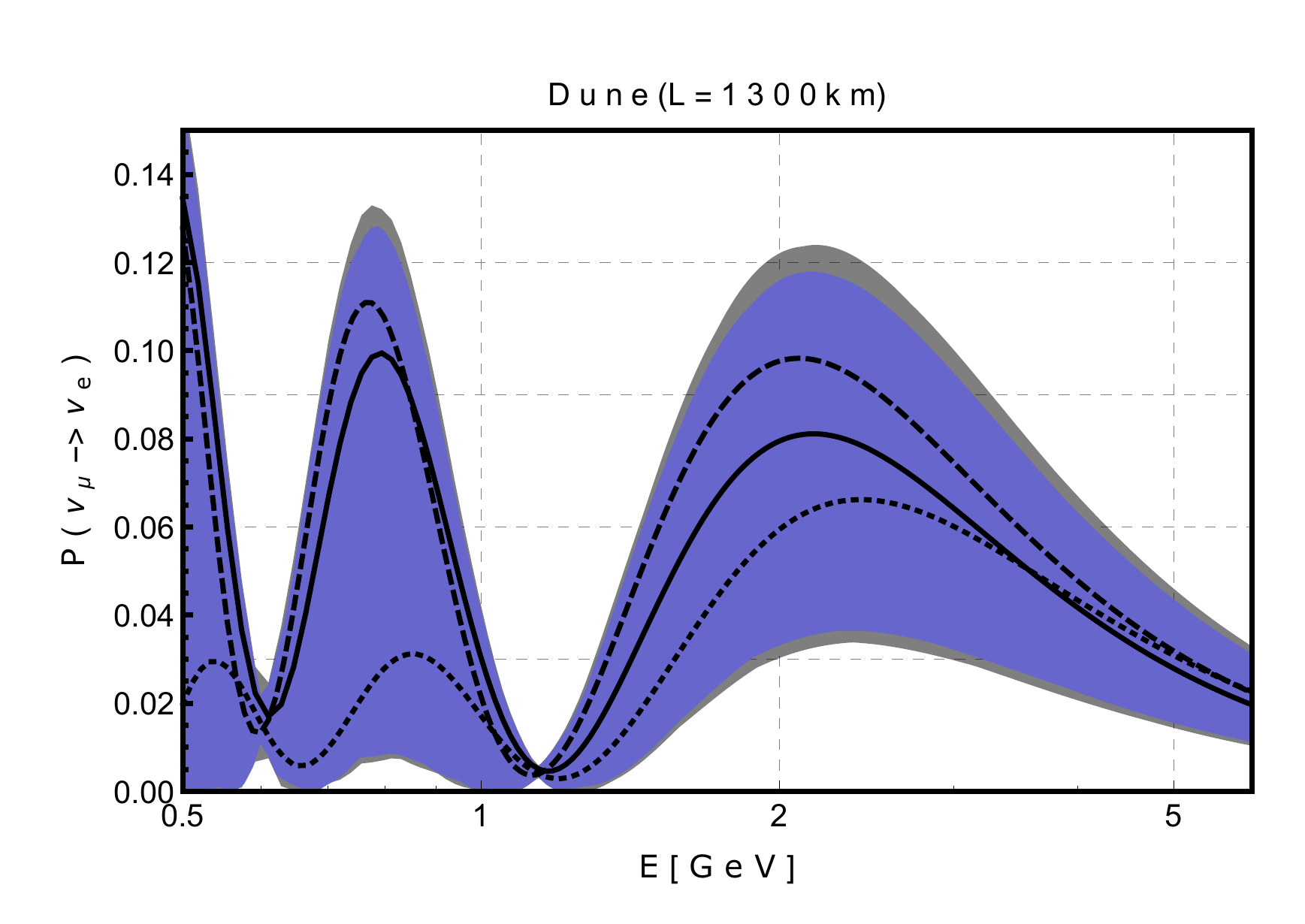}
    \includegraphics[scale=0.46]{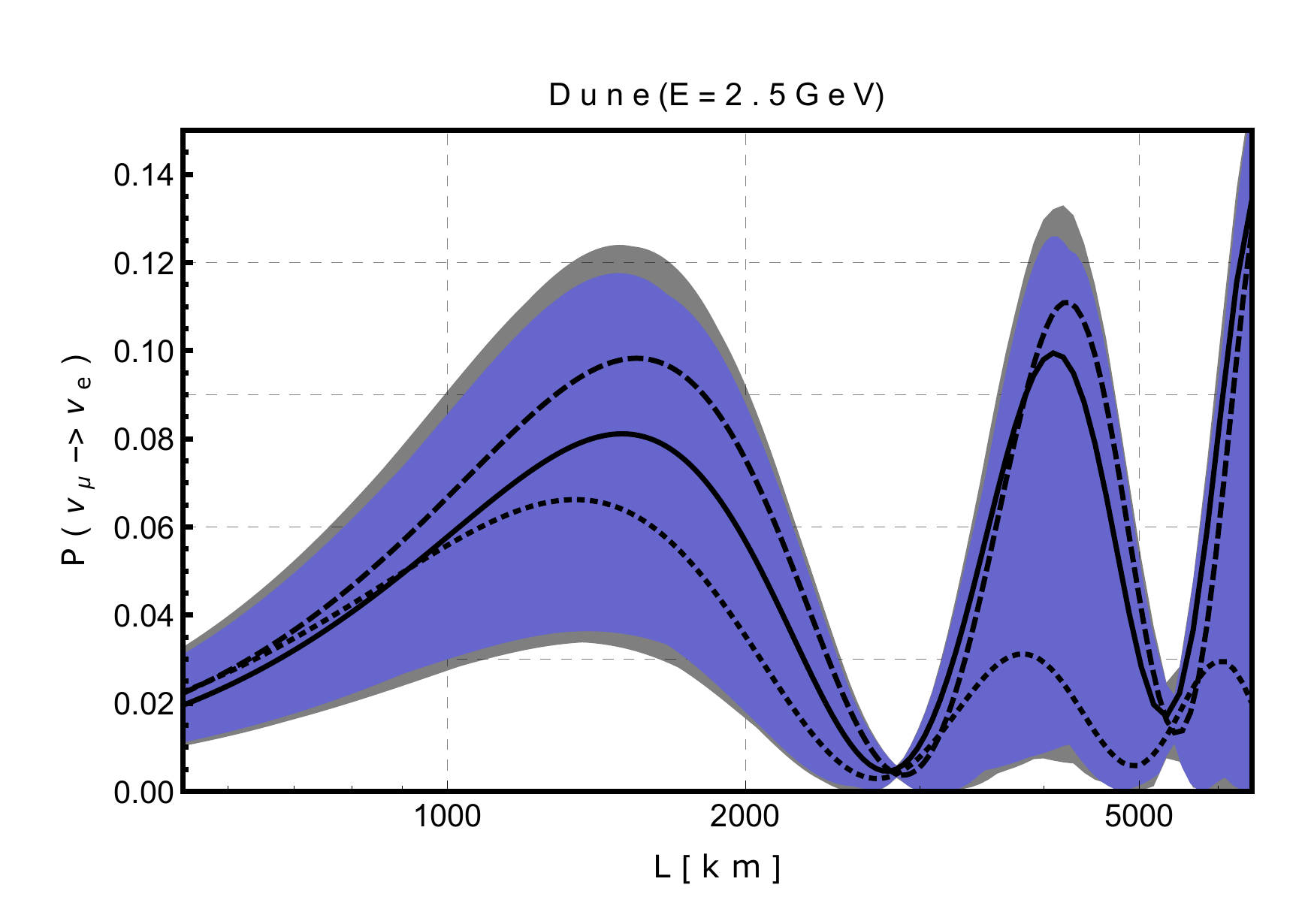}
    \caption{\label{fig:DUNE} {\bf Left:} Transition probability as a
      function of the neutrino energy, $E$, for the DUNE
      experiment. {\bf Right:} $P(\mu\to e)$ as a function of the
      baseline $L$ for the energy peak at $E=2.5 $ GeV. The solid line
      represents the current neutrino best fit with
      $\delta_{CP}=3\pi/2$, while the dashed and dotted lines
      correspond to minima with $\delta_{CP}=1.27\pi$ and with
      $\delta_{CP}=0.27\pi$, respectively.}
 \end{figure}

\begin{figure}[!h]
 \centering
    \includegraphics[scale=0.46]{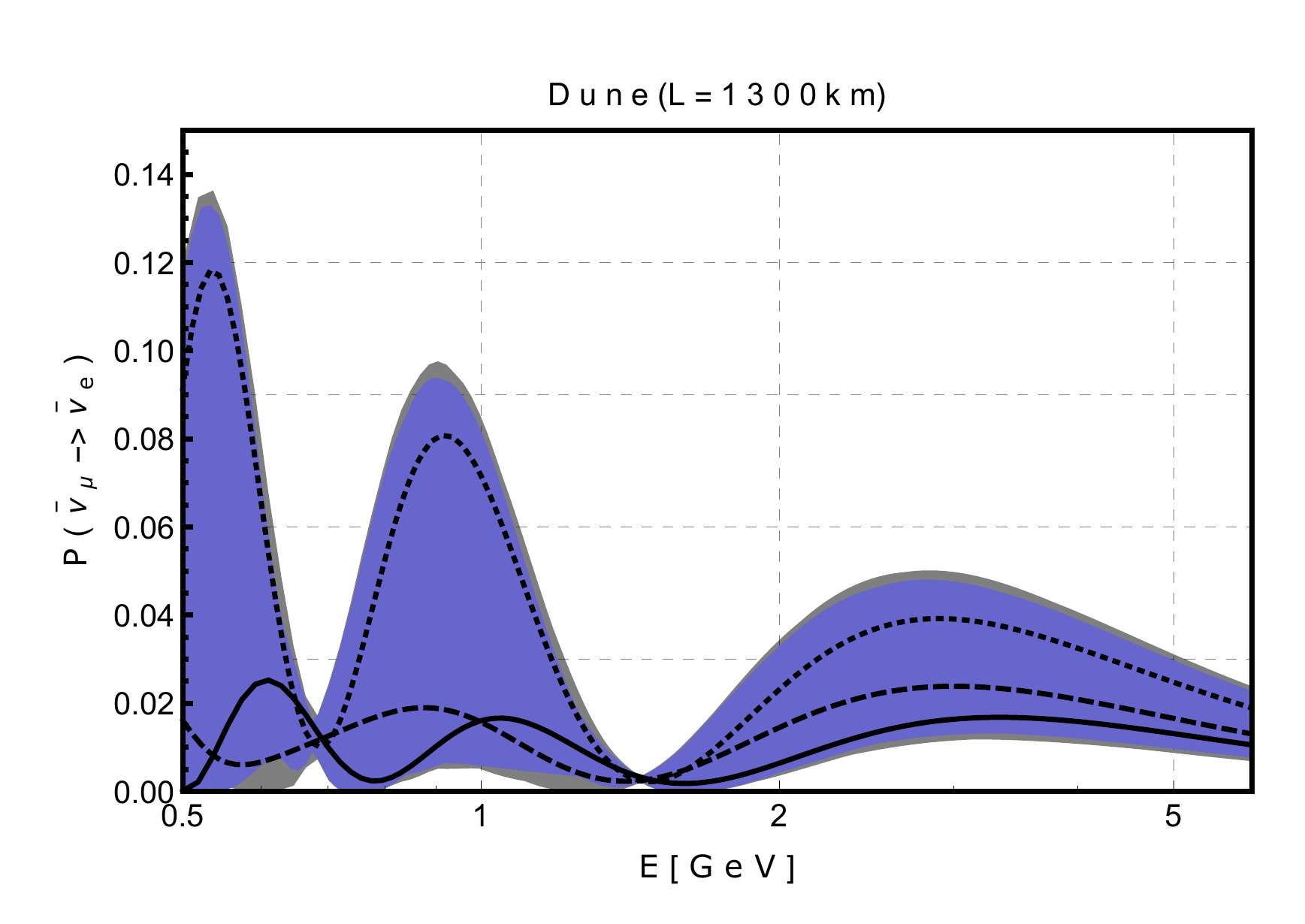}
    \includegraphics[scale=0.46]{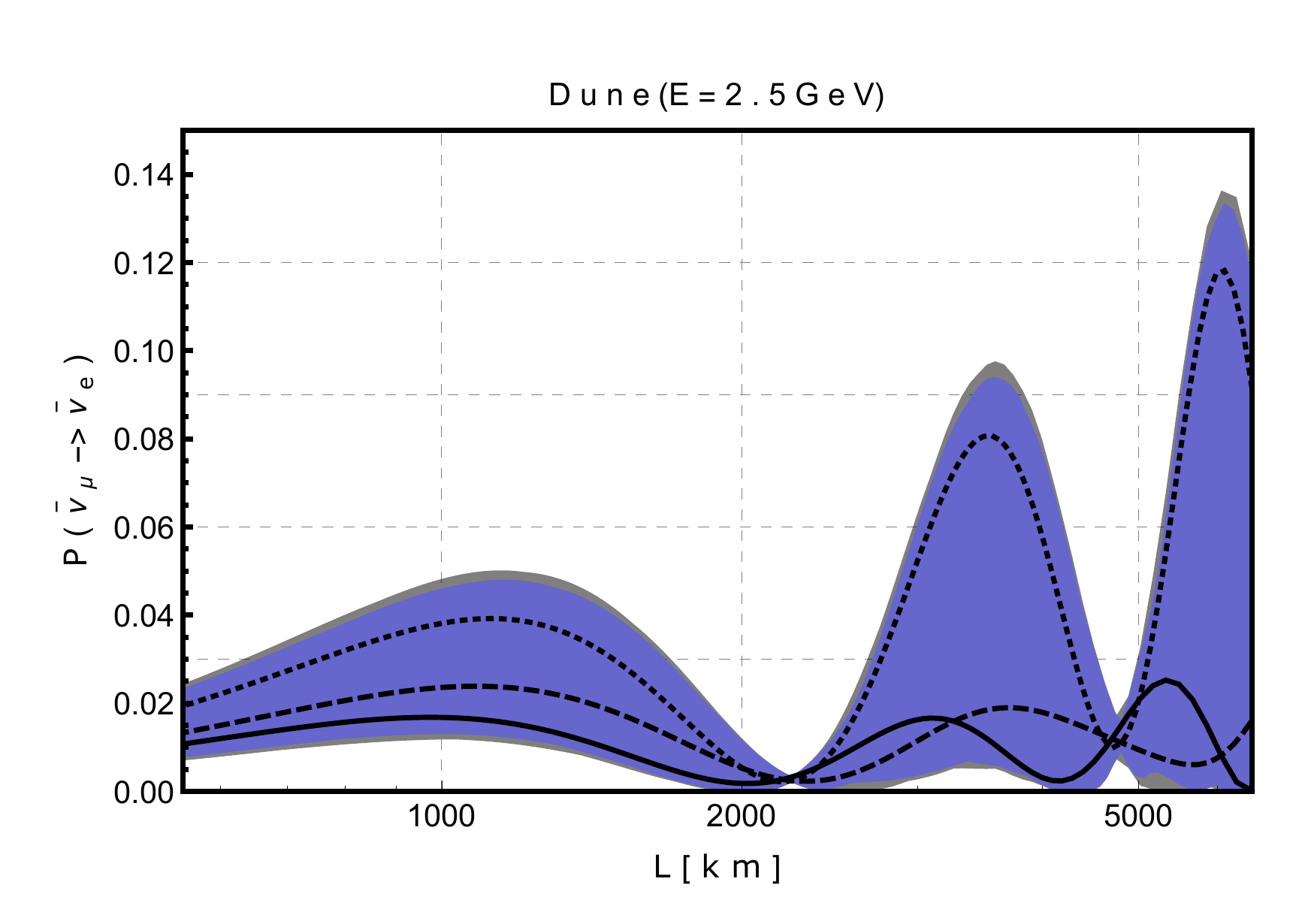}
    \caption{\label{fig:DUNE2} Same as Fig.~\ref{fig:DUNE}, but for
      the $\bar{\mu}\to \bar{e}$ oscillation probability.}
 \end{figure}

  In order to get a closer look on our model predictions compared to
  the general unconstrained case we display the neutrino oscillation
  probabilities expected in each experiment. These are plotted in
  Figs.~\ref{fig:T2K}-\ref{fig:DUNE2} for T2K, NO$\nu$A and DUNE
  respectively. The gray area corresponds to the case of no model
  constraints, while the color regions correspond to the predictions
  of our model. The solid line represents the unconstrained $\chi^2$
  minimum in each case with $\delta_{CP}=3\pi/2$, while the dashed and
  dotted lines correspond to minimum 1 and minimum 2 respectively, see
  Table. In Figs.~\ref{fig:T2K} and ~\ref{fig:Nova} the left panel
  gives the neutrino and the right one the anti-neutrino transition
  probabilities. Fig.~\ref{fig:DUNE} shows the same transition
  probability for neutrinos as a function of the energy in the left
  and as a function of the distance in the right, while
  Fig.~\ref{fig:DUNE2} shows the same for antineutrinos.

 From Figs.~\ref{fig:T2K} and \ref{fig:Nova} one sees that the
 currently allowed $3\sigma$ ranges for the oscillation probabilities
 are only slightly wider for the generic or ``unconstrained'' model
 than for our case, characterizing the good measurement of the mixing
 angles.
 The predictivity of our model consists mainly in the fourfold
 correlation depicted in the left panel in Fig.~\ref{fig:correlation}.
 The large difference between the dashed and dotted curves comes from
 the large difference in the $\delta_{CP}$ values associated to the
 minima in the Table which, in turn, are associated to the fourfold
 degeneracy. Indeed, as can be seen in Fig.\ref{fig:chi_delta}, these
 minima can have $\delta_{\rm CP}$ values very different from the
 current preferred one, especially in the first oscillation peaks.

 We now turn to the predictions for neutrino and anti-neutrino
 oscillation probabilities for the case of the DUNE experiment, with
 baseline 1300 km. In this case matter effects play a more important
 role than in the previous ones. As before we will take, for
 illustration, the deeper global minimum in Fig.~\ref{fig:chi_delta} (Minimum 1 with $\delta_{\rm CP}=1.27\pi$ )
 as the true one and compare it with Minimum 2 corresponding to
 $\delta_{\rm CP}=0.27\pi$ and with the ``standard'' case with
 $\delta_{\rm CP}=3\pi/2$.

 In Fig.~\ref{fig:DUNE} we show $\mu\to e$ the transition probability
 for DUNE as a function of the neutrino energy, $E$.  In the right
 panel we display the oscillation probability as a function of the
 baseline $L$ for peak energy at $E=2.5 $ GeV. The solid line
 corresponds to the current neutrino best fit with
 $\delta_{CP}=3\pi/2$, while the dashed and dotted ones correspond to
 the minima with $\delta_{CP}=1.27\pi$ and $\delta_{CP}=0.27\pi$,
 respectively. As an example one sees how the model allows for a
 potentially large shift in the position of the first oscillation
 maximum as well as its height, as seen in the expected shape of the
 transition probabilities, making the model potentially testable.

  \section{Discussion and Conclusion}

 We have characterized neutrino oscillation predictions emerging
 from a previously proposed warped model construction incorporating
 $\Delta(27)$ flavor symmetry~\cite{Chen:2015jta}.
 Fermion masses and mixing are all accommodated in a realistic way,
 while the pattern of neutrino oscillations is described in terms of
 two parameters $\cos\phi_\nu$ and $\sin2\theta_\nu$ nearly
 proportional to each other. The determination of $\delta_{CP}$ is
 fourfold degenerate with the minima allowed within 2$\sigma$, as seen
 in Fig.~\ref{fig:chi_delta}.
 The sharp correlation between $\delta_{CP}$ and the atmospheric
 mixing angle $\theta_{23}$ seen in Fig.~\ref{fig:correlation} encodes
 the true predictive power of the model, and implies that maximal
 $\theta_{23}$ is associated with maximal leptonic CP violation.
 We have seen how there can be substantial differences in the
 appearance neutrino and anti-neutrino probabilities for the long
 baseline accelerator experiments. Our results for the T2K, NO$\nu$A
 and DUNE experiments suggest that the model should potentially be put
 to a stringent test. Indeed these experiments could, in principle,
 rule it out, especially after improved determinations of
 $\delta_{CP}$ and the $\theta_{23}$ octant.

\section{Acknowledgements}

Work supported by Spanish grants FPA2014-58183-P, Multidark
CSD2009-00064, SEV-2014-0398 (MINECO) and PROMETEOII/2014/084
(Generalitat Valenciana).  P. S. P. acknowledges the support of
FAPESP  grant 2014/05133-1, 2015/16809-9 and 2014/19164-6. 

 \bibliographystyle{utphys}

 \providecommand{\href}[2]{#2}\begingroup\raggedright\endgroup

\end{document}